\documentclass[
aps,
pra,
longbibliography,
superscriptaddress,
 amsmath,amssymb,
twocolumn,
]{revtex4-1}

\usepackage[version=4]{mhchem} 

\usepackage{graphicx}
\usepackage{dcolumn}
\usepackage{bm}
\usepackage{upgreek}
\usepackage{color}
\usepackage{hyperref}
\DeclareMathAlphabet{\mathpzc}{OT1}{pzc}{m}{it}

\newcommand{\m}{\mathrm}

\newcommand{\eref}[1]{Eq.~(\ref{#1})}
\newcommand{\fref}[1]{Fig.~\ref{#1}}

\definecolor{blue}{RGB}{0,0,255}
\definecolor{red}{RGB}{255,0,0}
\definecolor{orange}{RGB}{200,100,0}

\usepackage[normalem]{ulem}

\makeatletter
\newcommand{\printfnsymbol}[1]{%
  \textsuperscript{\@fnsymbol{#1}}%
}
\makeatother

\makeatletter
\newcommand\thefontsize[1]{{#1 The current font size is: \f@size pt\par}}
\makeatother

\begin{document}

\title{Sideband control of a multimode quantum bulk acoustic system}

\author{Mikael Kervinen}
\thanks{Equal contribution}
\affiliation{Department of Applied Physics, Aalto University, P.O. Box 15100, FI-00076 AALTO, Finland}
\author{Alpo V\"alimaa}
\thanks{Equal contribution}
\affiliation{Department of Applied Physics, Aalto University, P.O. Box 15100, FI-00076 AALTO, Finland}
\author{Jhon E. Ram\'{i}rez-Mu\~noz}
\affiliation{Departamento de F\'{i}sica, Universidad Nacional de Colombia, 111321 Bogot\'{a}, Colombia}
\author{Mika A. Sillanp\"a\"a}
 \email{Mika.Sillanpaa@aalto.fi}
\affiliation{Department of Applied Physics, Aalto University, P.O. Box 15100, FI-00076 AALTO, Finland}%

\date{\today}

\begin{abstract}

Multimode bulk acoustic systems show promise for use in superconducting quantum computation. They can serve as a medium term memory storage, with exceptional coherence times demonstrated, and they exhibit a mode density that is physically highly compact. Here we experimentally demonstrate accessing individual acoustic modes without being hindered by the uniform frequency spacing of the modes. We use sideband control where a low-frequency modulation is applied to the transmon qubit energy. The amplitude of the modulation defines the qubit-acoustic mode coupling, and its frequency detunes the acoustic sidebands, therefore enabling selectively switching on or off the interaction, and allows for a full control on the individual modes.

\end{abstract}


\maketitle



\section{\label{sec:level1} Introduction}



Superconducting qubits are among the leading platforms for quantum computing. Besides the qubits themselves, quantum computation needs at least memory elements and interconnects. For that purpose, quantum hybrid systems \cite{Ashhab2013HybridReview,Kurizki3866} that combine superconducting qubits with disparate degrees of freedom have been actively investigated. Experimental work includes e.g.~systems containing spin or magnetic degrees of freedom \cite{Kubo2012SpinQB,Tabuchi2015}. 
A particularly actively explored medium for qubit hybrids are mechanical oscillators or acoustic modes in solid state \cite{LaHaye2009,OConnell2010,transmonnems,Delsing2014,Nakamura2017SAW,Leek2017SAW,Astafiev2018SAW,Cleland2019PhEntangl,Safavi2019Fock,Lehnert2019Fock}. A harmonic mode is a good candidate for a memory, and can also be used for continuous-variable quantum information \cite{Devoret2015CatCode}. Acoustic memory modes can be made very long-lived, as acoustic resonators with exceptionally long coherence times much beyond those of superconducting qubits have been demonstrated \cite{maccabe2019phononic}.

To maximize the benefit of integrating qubits with acoustic modes, the latter should have a compact form factor to be able to support multiple modes on-chip, in addition to the long coherence times. High overtone bulk acoustic wave resonators (HBAR) \cite{SchoelkopfHBAR2017,Kervinen2018,Chu2018,kervinen2019landau} can fulfil both of these criteria. They are naturally multimode systems and they can have exceptional quality factors. These systems can thus reduce the number of required qubits and control lines in order to create more hardware efficient systems \cite{Hann_2019}. 

In this work, we discuss and implement a novel approach to manipulate a qubit-HBAR system by controlling the qubit with longitudinal fields. The method has proven to be a useful tool to control superconducting quantum systems. The experimental demonstrations include sideband transitions \cite{strand2013first}, two qubit operations \cite{Caldwell_2018,Luyan2018ParMod}, a tunable coupler \cite{Gambetta2016ParMod}, all realised with flux induced parametric modulation.
%
%
The sideband transitions have also been used to generate a controllable interaction between a superconducting qubit and multiple microwave resonators \cite{Naik_2017}. 
Other experiments have used parametric modulation to generate photon-assisted Landau-Zener interference \cite{Silveri2015StuckelbergModulation, kervinen2019landau}, or motional averaging \cite{Li2013}.
Here, we show that the longitudinal modulation can be used to provide a selective access to the different acoustic modes in a harmonic multimode quantum acoustic system, where a transmon qubit is coupled to a HBAR device.

\section{Interaction using longitudinal modulation}



Let us treat a qubit with the transition frequency $\omega_0$ between the ground state and the first excited state. The qubit is described by the standard operators $\sigma_{z}$, $\sigma_{+}$ and $\sigma_{-}$. The qubit is coupled to an array of harmonic oscillators with frequencies $\omega_{m}^{(i)}$ for mode $i$, at the coupling energy $g_m$ that is supposed to be the same for all of the oscillators. The oscillators are described by the annihilation (creation) phonon operators $a_{i}$ ($a_{i}^{\dagger}$) for mode $i$.

\begin{figure}[!h]
   \includegraphics[scale=1]{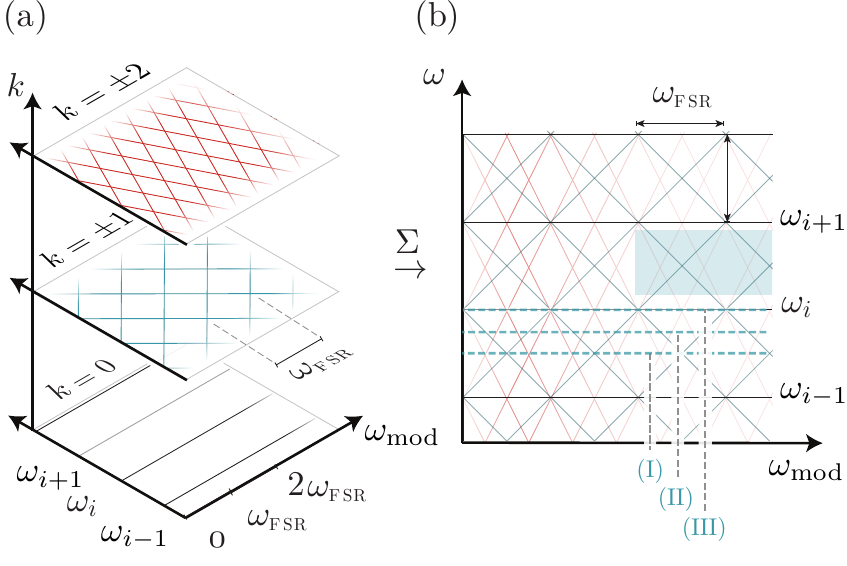}
    \caption{Graphical illustration of the acoustic sidebands. (a) Stacked representation of the acoustical sidebands from \eref{eq:H} as given by $\omega_{i}+k\omega_\text{mod}$. Each acoustical mode is subject to sidebands (shown up to $2^\m{nd}$ order) that linearly depend on the modulation frequency. 
    (b) Acoustic grids superimposed to represent a resonance landscape that we can probe with the qubit. The operating points (I-III) are defined with respect to the acoustic modes. The shaded area illustrates the measured parameter range in \fref{fig:time-domain_spectroscopy} (a). 
    }
    \label{fig:acoustic-grid}
\end{figure}

The qubit transition frequency is modulated longitudinally by a classical field $H_{z}(t)=\frac{A}{2}\cos{(\omega_\text{mod}t)}\sigma_{z}$. Here, $A$ is the amplitude of the modulation and $\omega_{\mathrm{mod}} \ll \omega_{0}$ is the modulation frequency. The full system under the frequency modulation is described by the Hamiltonian \cite{kervinen2019landau}
%
\begin{align}
\label{eq:H}
     H^{(n,k)}=&\frac{(n\omega_\text{mod}-\omega_0)}{2}\sigma_{z}+\sum_{i}(\omega_{i}+k\omega_\text{mod})a_{i}^{\dagger}a_{i} \nonumber \\
    &+\sum_{i}g_{m}J_{n-k}\left(\frac{A}{\omega_\text{mod}}\right)(\sigma_{+}a_{i}+\sigma_{-}a_{i}^{\dagger}) \,.
\end{align}
It describes $n^{\m{th}}$ order qubit sideband interacting at the rate $g_{m}J_{n-k}\left(A/\omega_\text{mod}\right)$ with $k^{\m{th}}$ order acoustic sideband. 
The modulation generates acoustic and qubit sidebands $(n,k = \pm 1, 2, ...)$ creating a grid of interweaving modes. In our analysis, we consider only the qubit main band $n=0$. Then, of relevance are the acoustic sidebands that are illustrated in \fref{fig:acoustic-grid}. 
%
%
%
By selecting the qubit's operation point suitably, different processes connecting the qubit and the acoustic modes can be realized via the frequency modulation. 
%
%
In \fref{fig:acoustic-grid}(b), the useful operating points are labelled (I)-(III). At point (I) without modulation, the qubit is decoupled from the acoustic modes and single qubit gates can be realized on the bare qubit. Similarly, at point (III) without modulation results in the normal resonant qubit-oscillator interaction. In order to access individual acoustic modes through the frequency modulation, the qubit is tuned to point (II) that is off from the mid point between consecutive acoustic mode frequencies. At this point, the sidebands become well separated in frequency and they can be accessed individually. At point (I), at exactly halfway between two acoustic modes, one can bring the qubit on resonance with two acoustic modes via the modulation.



\section{\label{sec:level2} Quantum bulk acoustic device}

\begin{figure}[!h]
   \includegraphics[width=0.48\textwidth]{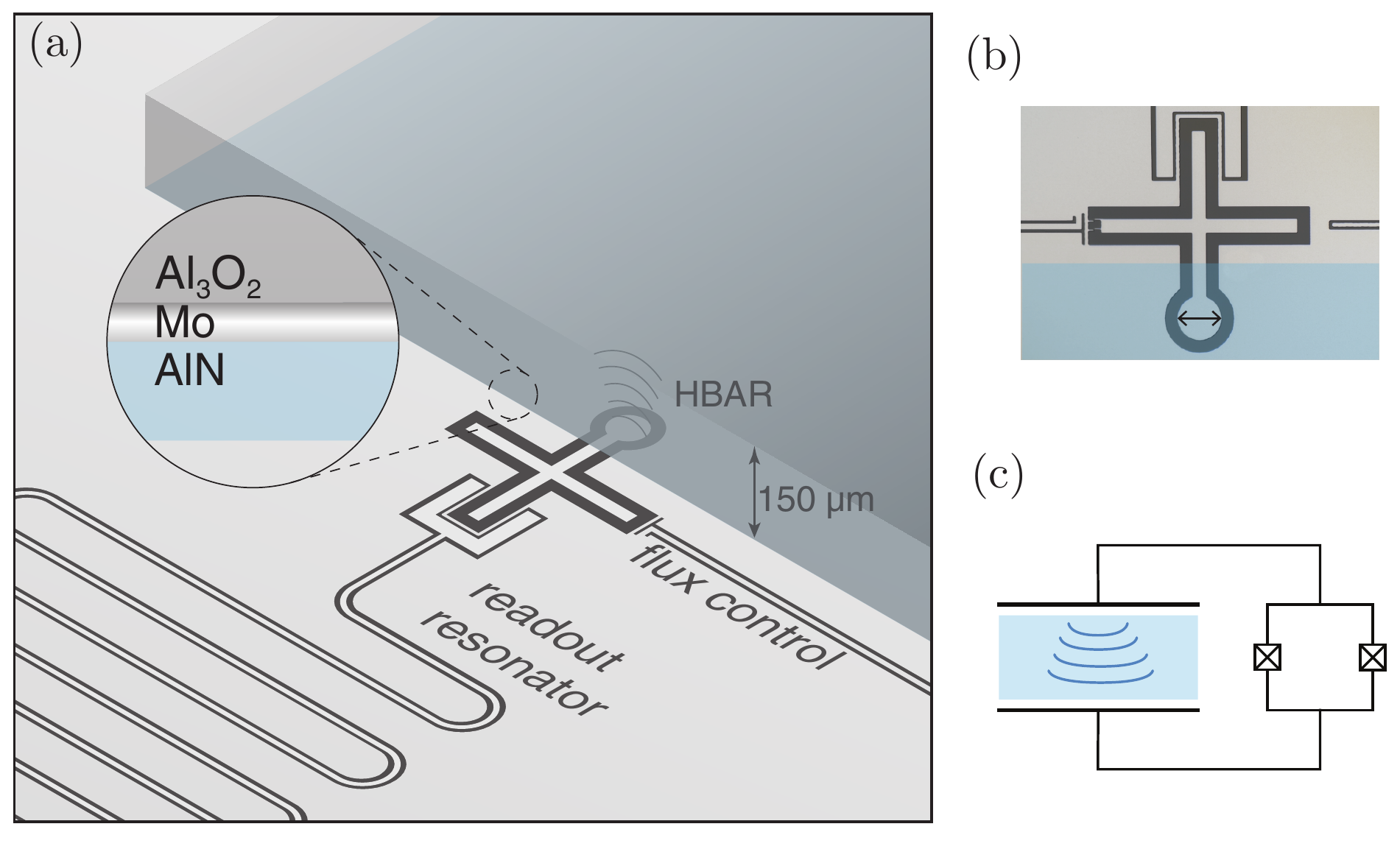}
    \caption{(a) Schematic of the sample, illustrating the flip-chip assembly of a high-overtone bulk acoustic wave resonator (blue) on top of the Xmon qubit. The acoustic medium is a sapphire crystal that is first covered by a thin layer of molybdenum (Mo, 60 nm), on top of which there is $\sim 1 \; \upmu$m thick layer of polycrystalline aluminium nitride.
    (b) The Xmon-qubit design has separate lines for flux control and transverse drive, a readout resonator with a coupling of $47$~MHz, and a circular pad for electro-acoustic coupling with the diameter of 72 $\upmu$m. (c) Schematic of the transmon-HBAR coupling. The piezoelectric layer acts as a transducer between the qubit's electric field and the acoustic modes.}
    \label{fig:sample}
\end{figure}

The experiments are realized with the sample depicted in \fref{fig:sample}(a). Compared to our previous device \cite{kervinen2019landau}, we have continued to improve the system parameters by redesigning the device geometry and the choice of materials. The device is implemented in a flip-chip design, where the high-overtone bulk acoustic resonator (HBAR) is overlayed on a a superconducting Xmon-type transmon qubit \cite{barends2013coherent} fabricated on a silicon substrate [\fref{fig:sample}(b)]. The qubit is capacitively coupled to the acoustic modes and the electroacoustic transduction is provided by a 900 nm thick polycrystalline aluminium nitride (AlN) and significantly enhanced by an underlying 60 nm-thick Molybdenum (Mo) layer. The HBAR substrate being 150 $\upmu$m thick single crystalline sapphire (\ce{Al3O2}) results in a free spectral range $\omega_\mathrm{FSR}/2\pi \approx 39$ MHz of the overtone modes. A thin AlN layer (30 nm) resides between the sapphire substrate and molybdenum to match the crystal lattices. The thin film stack has been manufactured by \textit{OEM Group}. The thickness of the film stack ensures the resonant conditions to favor electroacoustic coupling around 5.5 GHz. 

The (super)conducting Mo layer encloses the electrical fields above the Xmon coupling pad into a smaller volume to provide stronger electric fields that actuate the acoustic modes. In addition, the electric field is likely to be well uniform which reduces excess acoustic radiation into the bulk and onto the surface. In order to limit the dielectric dissipation in the potentially lossy piezoelectric layer and to mitigate the losses in the qubit \cite{wang2015surface}, the piezoelectric chip only partially covers the qubit. Therefore, the qubit's electric field participation ratio with the piezoelectric transduction is reduced. However, due to the enhanced coupling provided by the bottom electrode, this approach still retains a strong coupling, while maintaining a good qubit coherence. 


The HBAR chip is attached with a two component epoxy and the chip lies over the qubit without any spacers. We have observed that the natural gap between the two chips due to chip imperfections or small particles that land on the surfaces is on the order of a few hundred nanometers. In the current device, the piezoelectric layer is not shaped in any way which limits the lateral confinement of the acoustic modes that could result a lower acoustic quality factor as we have observed in these devices. Another contributing loss factor is the polycrystalline form of the AlN that may degrade the otherwise excellent lifetimes in the bulk acoustic resonator. Encouragingly, recent experimental advances have allowed to build completely epitaxial bulk acoustic wave resonators with a metallic bottom electrode creating highly coherent phonon modes  \cite{gokhale2020epitaxial}.

\begin{figure}[!h]
   \includegraphics[scale=1]{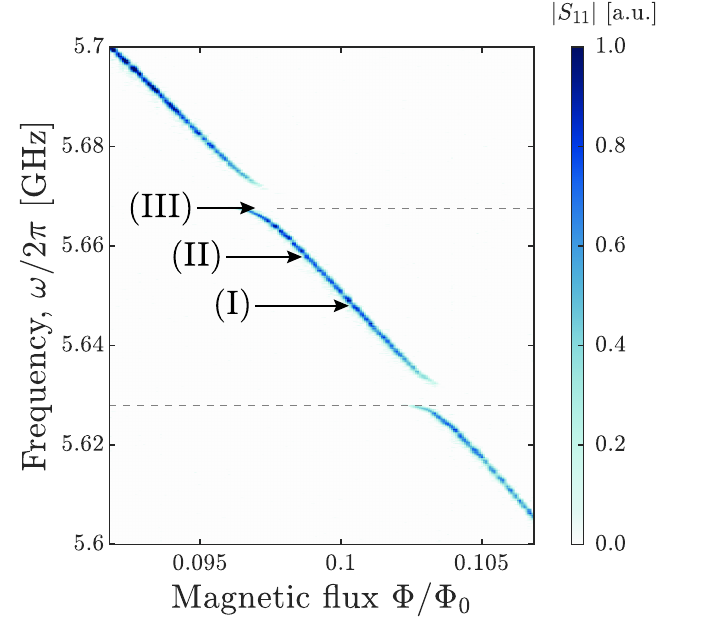}
    \caption{Two-tone spectroscopy as a function of flux bias. The operating points (I-III) correspond to those in \fref{fig:acoustic-grid}. DC-flux voltage may be applied to (I) decouple qubit from mechanics (without modulation), (III) bring the qubit on resonance with a single non-detuned acoustic mode (without modulation) or (II)  access individual acoustic sidebands (with modulation). Additionally, (I) with modulation is used to bring two acoustic modes simultaneously on resonance with the qubit.}
    \label{fig:sample-2tone}
\end{figure}

The Xmon qubit has symmetric junctions and a maximum frequency of 5.97 GHz. However, we operate the qubit away from the sweet spot and detune towards the enhanced qubit-mechanics coupling at 5.65 GHz, where the slope of the qubit energy curve is close to linear. The qubit is characterized in the operating point (I) [see \fref{fig:sample-2tone}], where the qubit is well decoupled from the acoustic modes owing to the detuning being much larger than the coupling. There we observe the qubit linewidth $\gamma/2\pi = 0.78$~MHz and measure its energy relaxation time $T_1 = 1.09 \; \upmu$s. By comparing to similar qubit samples without the flip-chip, we find that the presence of the piezo does not cause broadening of the qubit linewidth. Therefore, the piezoelectric film is not the current limiting factor of the qubit decoherence. We measure the acoustic lifetime by first exciting the qubit and then swapping the excitation to the mechanical mode. After a variable time delay the excitation is swapped back to the qubit and its population is measured. We find that the acoustic mode has a lifetime of $T_{1m} = 210$~ns corresponding to a decay rate of $ \gamma_{m}/2\pi = 0.76$~MHz. With the coupling rate of $g_m/2\pi \approx 3.1$~MHz, the system is in the strong coupling regime. 

\section{\label{sec:level3} Sideband control of the hybrid system}
\begin{figure}[!h]
  \begin{center}
   {\includegraphics[scale=1]{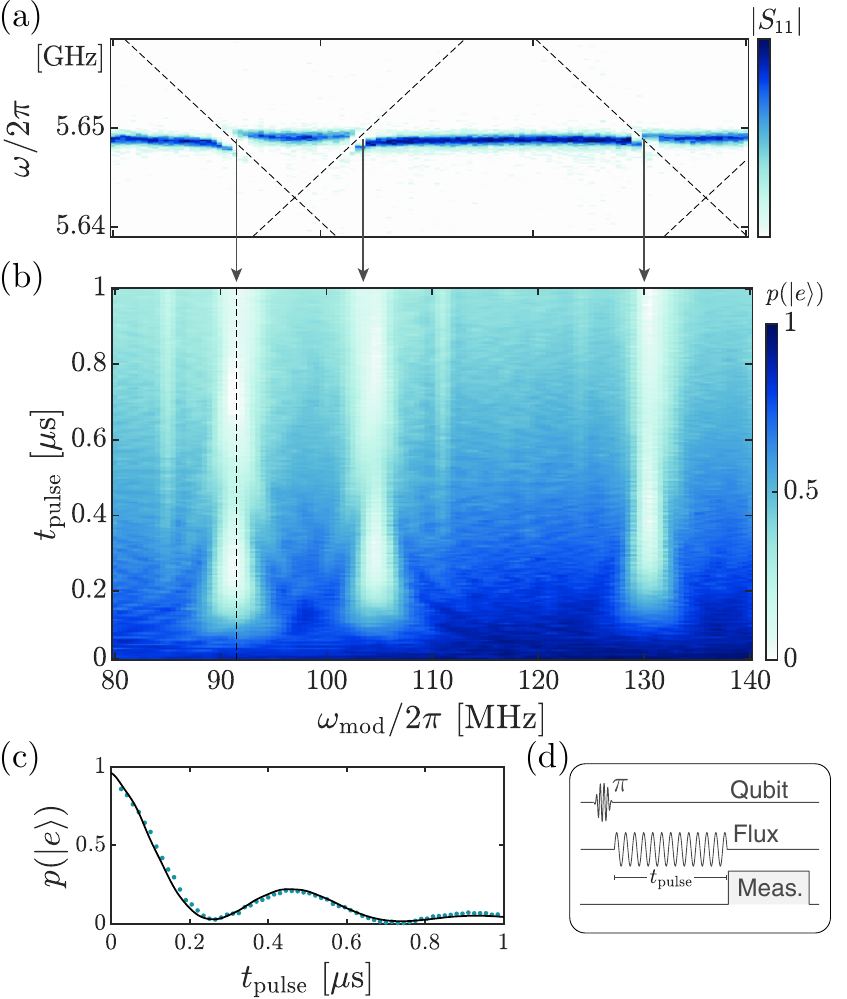}} 
    \caption{(a) Frequency-domain measurement to illustrate the operational point and the correspondence to the time-domain measurement and the related mechanical sidebands ($k= \pm 1$, dashed lines). (b) Time-domain spectroscopy. (c) A slice of the time-domain spectroscopy at the modulation frequency $\omega_{\m{mod}}/2\pi = 91$ MHz [indicated in (b)] together with theoretical modelling overlaid (solid line). (d) Schematic of the pulse sequence, where the qubit is first excited ($\pi$-pulse), followed by a modulation pulse and read out of the final qubit state. The length of the flux pulse is given by $t_{\m{pulse}}$. 
    }
    \label{fig:time-domain_spectroscopy}
 \end{center}
\end{figure}

We study the sideband interaction first in spectroscopic measurements. In \fref{fig:time-domain_spectroscopy}(a), the qubit is biased near the operation point (II), where individual acoustical sidebands are accessible. As the modulation frequency is varied from 80 MHz to 140 MHz, when the resonant condition $\omega_0 - \omega_i \pm \omega_\text{mod} = 0$ is fulfilled, we observe the different acoustic sidebands crossing the qubit frequency, at around $\omega_\text{mod}/2\pi \simeq 91, 105, 130$ MHz. The strength of each anticrossing is given by $2 g_{\m{eff}} = 2 g_m J_{1}(A/\omega_{\m{mod}})$, where $A/2\pi = 72$ MHz.

Next, we study these transitions in a time resolved manner. We operate the qubit at its resonace frequency. In the pulse sequence in \fref{fig:time-domain_spectroscopy}(d), the transmon is first excited with a $\pi$-pulse. We then apply the longitudinal flux modulation of a varying length and frequency. After the flux modulation pulse, the state of the transmon is measured with a measurement pulse. In \fref{fig:time-domain_spectroscopy}(b), we display the resulting data as a colormap showing the qubit excited state population. The horizontal axis is chosen to match that of panel (a). If the modulation frequency does not coincide with the difference between the qubit and an acoustic mode, we observe a relaxation decay of the bare qubit. Once the modulation frequency corresponds to the resonant condition with an acoustic mode, we observe coherent oscillations between the transmon and the given acoustic mode. These are photon-assisted vacuum Rabi oscillations where the oscillation rate is controlled by the amplitude of the modulation field according to the value of $g_{\m{eff}}$ under the given modulation condition. This allows coherent swapping of the qubit state with the state of a single acoustic mode. As seen in \fref{fig:time-domain_spectroscopy}(b), the photon-assisted vacuum Rabi oscillations of different acoustic modes are clearly separated allowing to selectively access individual modes. 


When the transmon population reaches its ground state, the excitation has been transferred to the acoustic mode. This is illustrated in detail in \fref{fig:time-domain_spectroscopy}(c), which displays a single time trace. Using the parametric modulation, the rate of the energy exchange is given by $2g_m J_{1}(A/\omega_{mod}) $. Here, the duration of the swap operation is about 240 ns. The fastest energy exchange that one can realize with the sidebands is $2g_m J_1(1.84) =  0.58 \times 2g_m$. However, at this point the coupling to the second order acoustic sidebands $k = \pm 2$ is also increased which could induce unwanted spurious interactions. These are faintly visible in \fref{fig:time-domain_spectroscopy}(b) at $\omega_\text{mod}/2\pi \simeq 85, 111, 123$~MHz. By a proper choice of the modulation amplitude and frequency, the higher order acoustic sidebands can be decoupled from the qubit [see Appendix \ref{sec:appendix}]. The modulation amplitude is selected such that the maximum peak of the sinusoidal modulation does not bring the qubit on resonance with an acoustic mode. However, due to the high mode density of the HBAR, the modulated qubit energy has to cross with the nearest acoustics modes which can cause increased decoherence.



\begin{figure}[!h]
  \begin{center}
   {\includegraphics{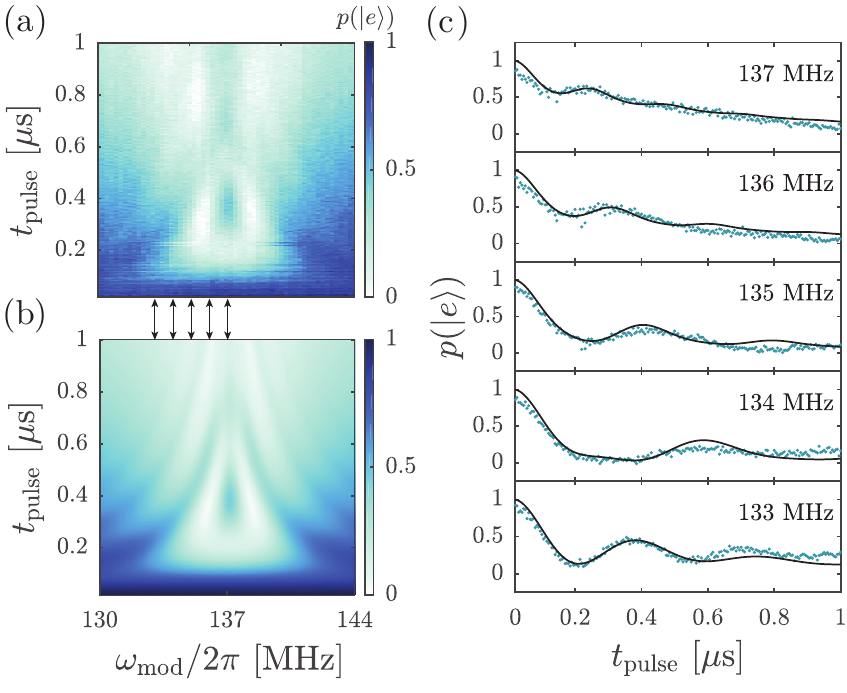}}
    \caption{(a) Experimental, and (b) theoretical time-domain study of two acoustic sidebands of the 1$^{\m{st}}$ order $k = \pm 1$ on resonance with the qubit. The qubit is first prepared in the excited state and then allowed to interact with the acoustic modes. (c) Single traces from (a), (b) at the modulation frequencies as indicated.}
    \label{fig:two_resonant_sidebands}
 \end{center}
\end{figure}

If the transmon is biased at the operation point (I) such that it is exactly in between two acoustic modes, the modulation creates acoustic sidebands with the two modes of opposite $k = \pm 1$. This can be used to couple the transmon to two acoustic modes simultaneously. The two acoustic modes are not necessarily the two closest modes but they can originate from several FSR away depending on the modulation frequency. When the qubit starts in its excited state and the two acoustic modes are brought into resonance by the modulation, the qubit excitation is split into the two acoustic modes creating entanglement of the tripartite system. We show the qubit population in \fref{fig:two_resonant_sidebands} at several detunings of the modulation frequency. 
 The two acoustic modes with opposite modulation frequency dependencies cross the qubit at 137 MHz, where we observe simultaneous vacuum Rabi oscillations between the qubit and both sideband modes.
 Compared to a case of a single resonant sideband, the oscillations experience faster energy exchange according to $2 g_\text{eff} \approx \sqrt{N-1}\times 2g_m J_{1}(A/\omega_{mod})$, where $N = 3$. When the modulation frequency coincides exactly with the frequency spacing to the two modes, the qubit is in resonance with the two acoustic modes simultaneously creating equal population in the acoustic modes.




\section{\label{sec:level4} Amplitude-controlled switch}

Next, we use the parametric flux modulation to control the resonant coupling between the qubit and an acoustic mode. We show how the flux modulation can be used to temporarily turn off the acoustic interaction. The use of longitudinal modulation to create a switchable coupling has been previously demonstrated with qubit-qubit and qubit-resonator systems \cite{wu2018efficient}.

\begin{figure}[!h]
  \begin{center}
   {\includegraphics{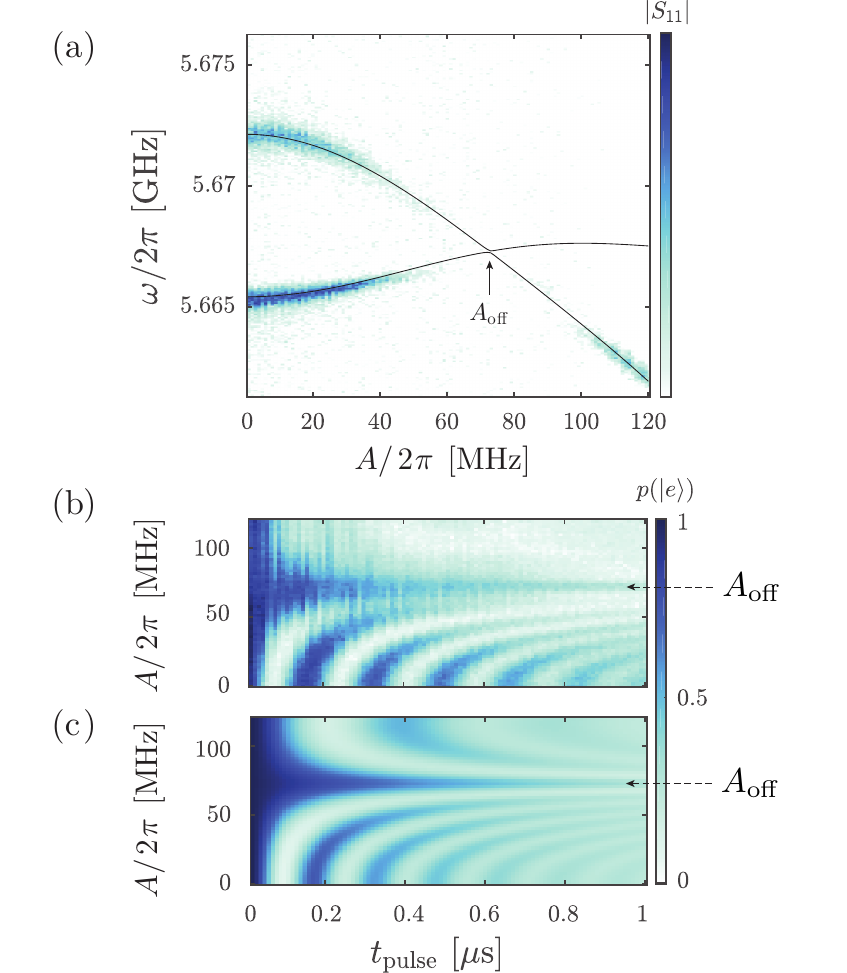}}
    \caption{Control of coupling by flux-modulation amplitude. 
    (a) 0$^\m{th}$ order acoustic mode on resonance with the qubit in a spectroscopic measurement and under the flux modulation. The frequency of the modulation is 30 MHz and the amplitude is varied across the root of the Bessel function that modulates the effective qubit-mechanics coupling. 
    (b) Time-domain spectroscopy of the qubit on resonance with an acoustic mode reveals the vacuum Rabi oscillations' frequency dependency on the flux modulation amplitude. The coupling, in essence, switches off at 72 MHz, where we see the decay of the qubit with no indication of the otherwise coupled acoustic mode. 
    (c) Theoretical modelling of a system comprising qubit and a single acoustic mode on resonance, to reproduce the measurement data in  (b).}
    \label{fig:amplitude-dependent_coupling}
 \end{center}
\end{figure} 

 A spectroscopic measurement of the modulated qubit-mechanics system shows an anticrossing in \fref{fig:amplitude-dependent_coupling}(a). The qubit frequency is fixed on resonance with an acoustic mode, and then the modulation is added on top of the static flux bias. As the amplitude is increased the effective coupling follows the Bessel function $J_0(A/\omega_\text{mod})$, where the modulation frequency is fixed at $\omega_\text{mod}/2\pi= 30$ MHz. Once the modulation amplitude-to-frequency ratio reaches $A/\omega_\text{mod} = 2.405$, that is the root of the zeroth order Bessel function, the interaction effectively vanishes demonstrating the possibility to switch off the coupling by selecting the appropriate amplitude-to-frequency ratio. The latter is clearly seen in \fref{fig:amplitude-dependent_coupling}(b),(c), where the vacuum Rabi oscillations freeze under the conditions of zero effective coupling. Due to a slight nonlinearity in the relationship between the DC-flux voltage and the qubit frequency, the driven AC flux induces rectification and contributes to the DC value in proportion with the driving amplitude, thus drifting the qubit away from resonance at higher amplitudes. This is clearly seen in \fref{fig:amplitude-dependent_coupling} (a) as the spectral lines bending down. The theoretical model (solid lines in \fref{fig:amplitude-dependent_coupling}(a)) matches well with the measurement.  



\begin{figure}[!h]
  \begin{center}
   {\includegraphics{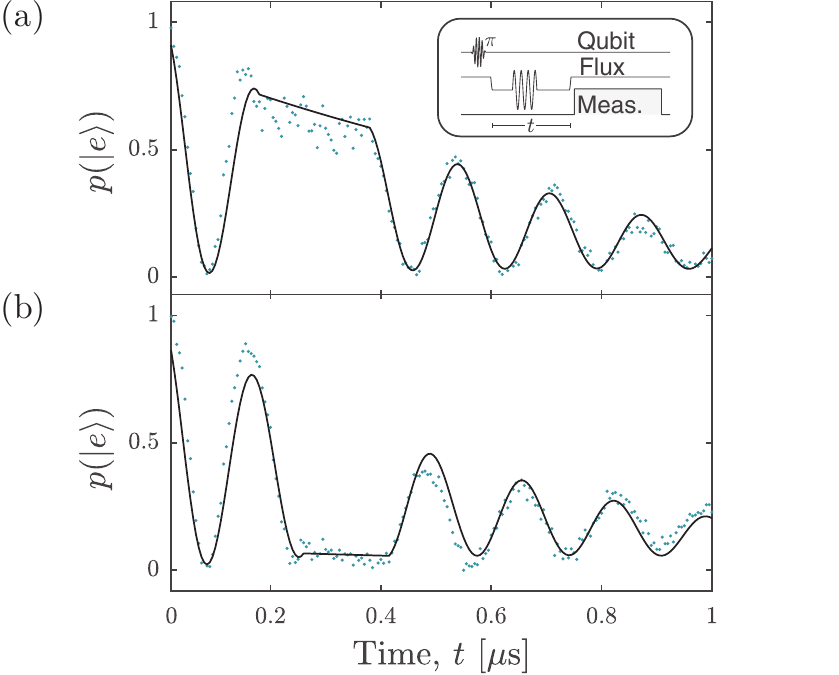}}
    \caption{Flux-modulation amplitude controlled switch. 
        The interaction is switched off, when the quantum is (a) in the qubit or (b) in the acoustic mode, after which the interaction is recovered. Solid black line is from a master equation simulation. The inset in (a) shows the pulse sequence consisting of a $\pi$-pulse to prepare the qubit in the excited state, a DC-flux shift to bring the qubit on resonance with the acoustic mode, flux modulation at appropriate amplitude and frequency to switch off the interaction, and the qubit readout.}
    \label{fig:switch_on-off}
 \end{center}
\end{figure} 

The parametric flux modulation can also be used to control the resonant coupling between the qubit and an acoustic mode on-the-fly during a single time-domain experiment (see \fref{fig:switch_on-off}). In this case, the modulation pulse is timed as in \fref{fig:switch_on-off} (a), where the time-dependent modulation parameters yield time-varying qubit-mechanics coupling and interruption of vacuum Rabi oscillations. When the amplitude of the modulation field is zero, the system experiences the regular vacuum Rabi oscillations between the qubit and the acoustic resonator at a rate of 6 MHz. The application of the switch is demonstrated in \fref{fig:switch_on-off} (a) and (b). The vacuum Rabi oscillations are allowed to evolve for $\sim 200$ ns until a switch-off pulse is applied when the quantum is either in the qubit or in the acoustic mode. 

After the switch off pulse has ended, the system resumes the coherent oscillations but with a reduced amplitude due to the energy decay. This shows an example of the utilization of the acoustic resonator as a memory element, where the quantum state is stored inside the acoustic mode and retrieved back to the qubit. With improved mechanical lifetimes, one could expect usage as a medium-term memory.  





\section{\label{sec:level6} Conclusions}
The control techniques that we have demonstrated in this work are based on the longitudinal modulation of the qubit energy level which creates a tunable interaction with the acoustic modes. Although the acoustic modes are equidistant from each other in frequency, by a suitable choice of the parameters one can accurately access individual modes, and also bring two acoustic modes on resonance.

We have also demonstrated a new design for a quantum acoustic system that combines a strong coupling to the acoustic modes with a high qubit coherence. This is achieved by reducing the qubit's electric field participation ratio with the piezoelectric transduction while enhancing the electromechanical coupling with the usage of a bottom electrode in the HBAR film stack. The device parameters can be greatly improved by combining state-of-the-art qubits with highly coherent acoustic modes. 

\appendix

\begin{figure}[!h]
  \begin{center}
{\includegraphics[scale=1]{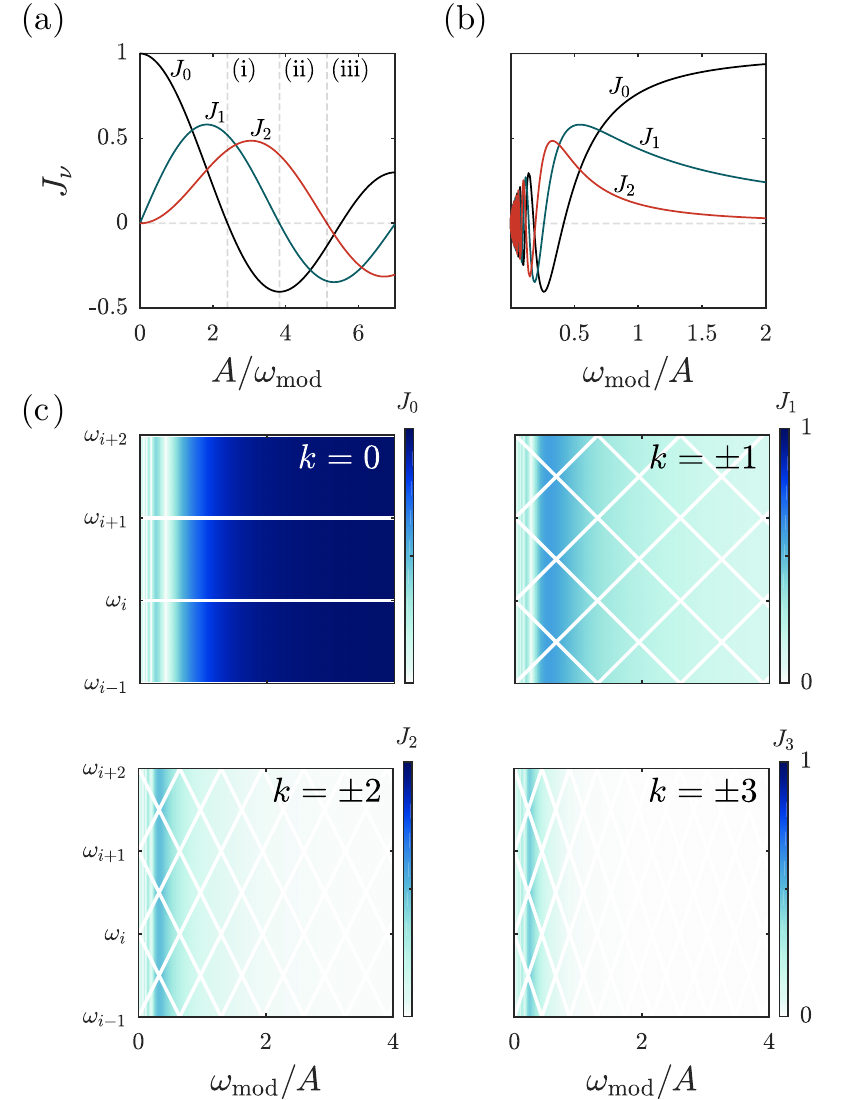}} 
    \caption{(a) Bessel functions of the first kind traced shown for the lowest three orders $J_0$, $J_1$ and $J_2$. Positions (i-iii) mark the zeros of the functions. (b) With the inverse argument, the same functions describe the $(A,\omega_\m{\textrm{mod}})$-dependence of the effective coupling, $g_{\textrm{eff}}$. (c) The colormap visualizes the value of $g_{\textrm{eff}}$ for each sideband order $|k|$ and $\omega_{\textrm{mod}}/A$ between 0 and 4. The overlaid white lines illustrate the acoustic sidebands.  
    }
    \label{fig:bessel}
 \end{center}
\end{figure} 

\section{Effective coupling}
\label{sec:appendix}
In \fref{fig:acoustic-grid}, we display the underlying network of acoustic sidebands. However, the coupling to the sidebands is introduced by the modulation parameters ($A$, $\omega_\m{mod}$). As discussed, by selecting a suitable operating point, one can find regions where the individual acoustical sidebands are accessible. Then, 
one suppresses the coupling to the higher-order acoustic sidebands via the factor $J_{n-k}(A/\omega_\m{mod})$. With a sufficiently small modulation amplitude, or high enough modulation frequency, the higher order acoustic sidebands loose a significant coupling to qubit. This is because the Bessel functions of the first kind $J_{n-k}$ approach zero, when $A/\omega_\m{mod}$ approaches zero at an increasing rate with the increasing order of the Bessel functions. Indeed, one can find that $J_{n-k}(A/\omega_\m{mod}) \propto (A/\omega_\m{mod})^{n-k}$ at small arguments. In \fref{fig:bessel}(a) we display the Bessel functions for $J_0$, $J_1$ and $J_2$. At the root of the functions one can find a total suppression of the coupling. The inverse argument $\omega_\m{mod}/A$ that relates to \fref{fig:time-domain_spectroscopy}(b) is illustrated in \fref{fig:bessel}(b). The choice of parameters $(A,\omega_\m{\textrm{mod}})$ is guided by a sufficiently strong coupling $g_mJ_1$ and a reasonable suppression of higher order terms. Beyond $\omega_\m{mod}/A>0.5$, the coupling to the higher order acoustic sidebands is quickly diminished. The sideband energies of the system are mapped to the coupling in \fref{fig:bessel}(c), with the colormap displaying the effective coupling with fixed $\omega_\text{mod}/\omega_\text{FSR}$ ratio.




\begin{acknowledgments} This work was supported by the Academy of Finland (contracts 308290, 307757, 312057), by the European Research Council (contract 615755), by the Centre for Quantum Engineering at Aalto University, by The Finnish Foundation for Technology Promotion, and by the Wihuri Foundation. We acknowledge funding from the European Union's Horizon 2020 research and innovation program under grant agreement No.~732894 (FETPRO HOT). We acknowledge the facilities and technical support of Otaniemi research infrastructure for Micro and Nanotechnologies (OtaNano) that is part of the European Microkelvin Platform. We would like to thank Wayne Crump for useful discussions.
\end{acknowledgments}


%

\end{document}